\documentclass[prd, aps, superscriptaddress, preprintnumbers, twocolumn, floatfix, nofootinbib]{revtex4}
\pdfoutput=1

\usepackage{amsfonts}
\usepackage{amsmath}
\usepackage{amssymb}
\usepackage{bm}
\usepackage{dcolumn}
\usepackage{graphicx}   
\usepackage[latin1]{inputenc}
\usepackage{latexsym}
\usepackage{rotating}
\usepackage{hyperref}
\usepackage{graphicx}
\usepackage{color}

\newcommand\be{\begin{equation}}
\newcommand\ba{\begin{eqnarray}}
\newcommand\ee{\end{equation}}
\newcommand\ea{\end{eqnarray}}

\newcommand{\mat}[1]{\mbox{\boldmath{$#1$}}}

\begin{document}

\title{A Superfluid Dark Matter Cosmic String Wake}

\author{Aline Favero}
\email{aline.favero@mail.mcgill.ca}
\affiliation{Department of Physics, McGill University, Montr\'{e}al, QC, H3A 2T8, Canada}

\author{Robert Brandenberger}
\email{rhb@physics.mcgill.ca}
\affiliation{Department of Physics, McGill University, Montr\'{e}al, QC, H3A 2T8, Canada}

\date{\today}

%%%%%%%%%%%%%%%%%%%%%%%%%%%%%%%%%%%%%%%%%%%%%%%%%%%%%%%%%%%%%%%%%%%%%%%%%%%%%%%%%%%%%%%%%%%%%%
\begin{abstract}

We study the effects of superfluid dark matter on the structure of a cosmic string wake,  considering both the effects of regular and quantum pressure terms.  We consider the total fluid to consist of a combination of baryons and dark matter. Hence, we are also able to study the effects of superfluid dark matter on the distribution of baryons inside the wake. We focus on parameter values for the superfluid dark matter which allow a MONDian explanation of galaxy rotation curves.

\end{abstract}
%%%%%%%%%%%%%%%%%%%%%%%%%%%%%%%%%%%%%%%%%%%%%%%%%%%%%%%%%%%%%%%%%%%%%%%%%%%%%%%%%%%%%%%%%%%%%%

\maketitle

%%%%%%%%%%%%%%%%%%%%%%%%%%%%%%%%%%%%%%%%%%%%%%%%%%%%%%%%%%%%%%%%%%%%%%%%%%%%%%%%%%%%%%%%%%%%%%
\section{Introduction} 
\label{sec:intro}

Wakes \cite{wake} formed behind moving long cosmic strings are responsible for some of the key observational signatures of strings (see e.g. \cite{CSrevs} for reviews of the cosmology of cosmic strings). Wake formation is usually studied assuming that the dark matter is completely cold (see, however, the study of \cite{HDMwake} for wake formation if the dark matter is hot, and \cite{Sorn} for a study of the effects of baryons on the formation of a cold dark matter wake).  Recently, however, there has been increasing interest in the possibility that the dark matter might be a superfluid \cite{Khoury} since in this case the empirical success of MOdified Newtonian Dynamics (MOND) \cite{MOND} on galactic scales can be reconciled with the successes of cold dark matter on cosmological scales. In this context the superfluid may also yield a candidate for Dark Energy and hence provide a unified dark sector model \cite{Elisa}. 

In this paper we study the accretion of superfluid dark matter and baryons onto a cosmic string wake. We are specifically interested in how both the regular and the quantum pressure of the superfluid change the properties of the string wake which then affect the resulting observational signatures\footnote{In a previous paper \cite{Previous}, we have studied whether a string can trigger superfluid condensation when moving through a fluid which is not yet condensed, and we also studied superfluid shock formation triggered by the wake.}.  As an interesting example, we can consider values for the superfluid which allow a MONDian explanation of galaxy rotation curves. This is reflected in the particular form of the dark matter pressure term (see below).

There  is a good reason from particle physics (see e.g. \cite{RHBrev2} for a review) to study cosmic strings. String solutions exist in a subclass of models beyond the Standard Model of particle physics. If Nature is described by such a model, then a causality argument due to Kibble \cite{Kibble} implies that a network of strings will form in a symmetry breaking phase transition in the early universe and persist to the present time.  Cosmic strings are lines of trapped energy, and their gravitational effects lead to signatures in many windows to probe the universe.  Many cosmological effects of cosmic strings are proportional to their tension $\mu$ which in turn is proportional to $\eta^2$, where $\eta$ is the energy scale of the particle physics phase transition which leads to cosmic string production. Hence, searching for cosmological signatures of strings is a way to probe particle physics ``from top down'' in the sense that the string signatures are larger if the energy scale of the new physics is higher.  

\begin{figure}[t]
	\centering
	\includegraphics[scale=.5]{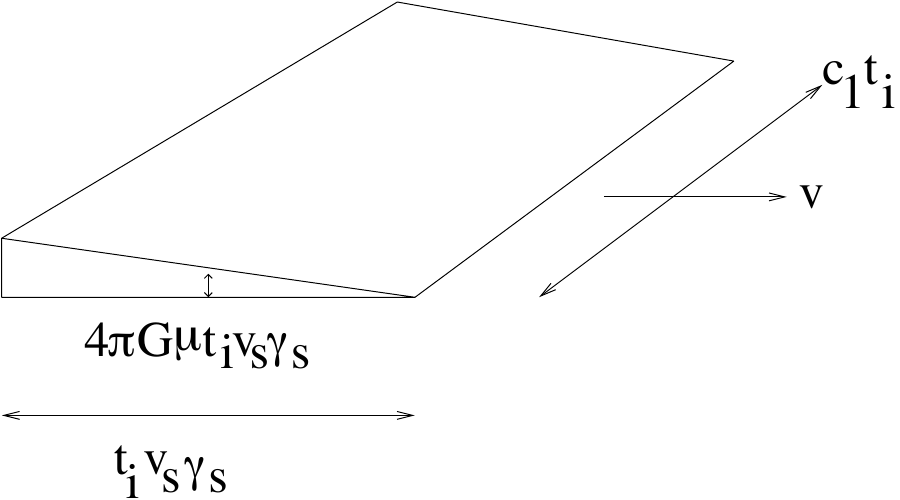}
	\caption{Sketch of a cosmic string wake produced at time $t_i$ by a string segment of length $c_1 t_i$ (where $c_1$ is a constant of order unity) and moving with velocity $v = v_s$ in the transverse direction. The depth of the wake is given by the Hubble time multiplied by $v_s \gamma_s$, where $\gamma_s$ is the relativistic gamma factor associated with $v_s$.}  
	\label{fig00}
\end{figure}

Typically, the network of strings approaches a dynamical fixed point, a ``scaling solution'' in which the statistical properties of the network of strings are independent of time if all lengths are scaled to the Hubble radius.  In particular, this implies that at all times each Hubble volume will be crossed by a time-independent number $N$ of ``long'' strings \cite{onescale}.  Numerical simulations \cite{CSsimuls} indicate that $N \sim 10$ (this is an order of magnitude statement).  Since strings are relativistic objects, their transverse velocity is typically of the order of the speed of light $c$.  Since space perpendicular to a long string is conical with a deficit angle $8 \pi G \mu$ (where $G$ is Newton's gravitational constant), a long straight string segment moving with velocity $v_s$ in transverse direction will produce a velocity fluctuation towards the plane behind the moving string\footnote{The plane is spanned by the vector tangent to the string and the velocity vector.}.  This will induce a {\it wake}, an overdense region behind the string \cite{wake}. Fig.~\eqref{fig00} represents a sketch of a string wake in the case that the matter is idealized cold dark matter (no pressure). 

A string segment present at time $t_i$ will produce a wake whose initial physical extent is $c_1 t_i$ in direction of the string ($c_1$ being a constant of order one), $v_s \gamma_s t_i$ in direction of motion of the string (with $\gamma_s$ being the relativistic gamma factor associated with the velocity $v$), and with mean initial thickness $4 \pi G \mu v_s \gamma_s t_i$. Once formed, the wake will maintain constant comoving extent in the planar directions, while the comoving thickness will grow through gravitational accretion. This growth is usually studied in the Zel'dovich approximation \cite{Zel}. In this paper we will study the wake growth if the dark matter is a superfluid. We will also take into account the presence of baryons, and consider both the size of the dark matter and baryon wakes.

We make use of natural units of which $c = \hbar = k_B = 1$, and work in the context of a homogeneous, isotropic and spatially flat cosmology with scale factor $a(t)$ normalized to be $a(t_0) = 1$ at the present time $t_0$.  The derivative with respect to time $t$ is indicated by an overdot. The Planck mass is denoted as $M_{Pl}$, and $G$ is Newton's gravitational constant. Since the dominant wake signals are from wakes formed in the phase of matter domination, we will focus on this phase and take $a(t) \sim t^{2/3}$ for $t > t_{\mathrm{eq}}$,  where $t_{\mathrm{eq}}$ is the time of equal matter and radiation.

\section{Accretion of Superfluid Dark Matter onto a Cosmic String Wake}
\label{sec2}

In this section we derive the equations of motion for the accretion of superfluid dark matter and baryons onto a string wake,  assuming planar symmetry. We work in the context of the Zel'dovich approximation \cite{Zel} and thus consider the evolution of mass shells which have a physical distance $\mathbf{r}$ from the wake. We can take a set of Cartesian coordinates in which the cosmic string is laying along the $z$-axis and moving in the negative $y$-direction, so that its wake is located in the $(x, y)$ plane. We consider a wake produced by a string segment at time $t_i$ (with $t_i > t_{\mathrm{eq}}$).

The physical heights of shells of superfluid dark matter and baryons, respectively, can be written as
\begin{align} \label{exp}
	\mathbf{r}_{\mathrm{DM}}\left(\mathbf{x}_{\mathrm{DM}}, t\right) &= a\left(t\right) \left[\mathbf{x}_{\mathrm{DM}} + \mat{\psi}_{\mathrm{DM}}\left(\mathbf{x}_{\mathrm{DM}}, t\right)\right],\\
	\mathbf{r}_{b}\left(\mathbf{x}_{b}, t\right) &= a\left(t\right) \left[\mathbf{x}_{b} + \mat{\psi}_{b}\left(\mathbf{x}_{b}, t\right)\right], \nonumber
\end{align}
subject to the initial conditions 
\ba \label{IC}
\mat{\psi}_{\mathrm{DM}}\left(\mathbf{x}_{\mathrm{DM}}, t_{i}\right) &=& 0 = \mat{\psi}_{b}\left(\mathbf{x}_{b}, t_{i}\right)\, , \\
\dot{\mat{\psi}}_{\mathrm{DM}}\left(\mathbf{x}_{\mathrm{DM}}, t_{i}\right) &=& - \mathbf{u}_{i} \epsilon\left(\mathbf{x}\right) a^{-1}\left(t_{i}\right), \,\,\, {\rm{and}} \nonumber \\ \dot{\mat{\psi}}_{b}\left(\mathbf{x}_{b}, t_{i}\right) &=& - \mathbf{u}_{i} \epsilon\left(\mathbf{x}\right) a^{-1}\left(t_{i}\right), \nonumber
\ea
where 
\be
\mathbf{u}_{i} = 4 \pi G \mu v_{s} \gamma_{s} \hat{\mathbf{x}}
\ee
is the initial velocity kick given by the string wake to the particles, and
\begin{align}
	\epsilon\left(\mathbf{x}\right) = \theta\left(\mathbf{x}\right) - \theta\left(- \mathbf{x}\right) = \begin{cases}
		+ 1, \quad \mathrm{for} \;\mathbf{x} > 0,\\
		- 1, \quad \mathrm{for} \;\mathbf{x} < 0.
	\end{cases}
\end{align}
Here, $\mathbf{x}_{\mathrm{DM}}$ and $\mathbf{x}_b$ denote the initial comoving distances of the dark matter and baryon shells from the wake, and $\mat{\psi}_{\mathrm{DM}}$ and $\mat{\psi}_b$ are the induced comoving displacements due to the initial velocity towards the wake and the gravitational accretion by the wake overdensity. 

Note that the time evolution of the comoving displacements depends on the shell which we are considering, i.e. on the value of $\mathbf{x}$. The width of the wake at time $t$ is determined by the shell which is ``turning around'' at time $t$, i.e.  by the value of $\mathbf{x}$ for which ${\dot{\mathbf{r}}} = 0$. This condition applied to the dark matter shells yields the width of the dark matter wake while the same condition applied to baryonic shells yields the width of the baryonic wake.  From (\ref{exp}) it follows that the comoving width of a wake of type $a$ (where $a$ stands either for dark matter or for baryons) is given by
\be \label{width}
\mathbf{x}_a(t) \, = \, - \mat{\psi}_a(t) - H^{-1}(t) {\dot{\mat{\psi}}_a(t)} \, .
\ee
 It is often assumed that after turnaround matter will virialize at half of this distance. In this paper we are concerned with the differences in the wake sizes if the dark matter is a superfluid compared to if the dark matter is pressureless. Hence, we will be focusing on differences in the values of the displacements.
 
The Zel'dovich approximation is based on Newtonian gravity
\be \label{Newton}
{\ddot{r}}^{i} \, = \, - \frac{\partial \phi}{\partial r^{i}},
\ee
where $\phi$ is the Newtonian gravitational potential which is determined by the Poisson equation
\be
\nabla^2 \phi = 4 \pi G \rho \, ,
\ee
$\rho$ being the energy density.  Pressure forces lead to an extra term on the right hand side of (\ref{Newton}).  For the planar geometry which we are assuming the gradient reduces to the partial derivative in the $x$-direction.

In the absence of non-gravitational coupling between baryons and dark matter, dark matter and baryons both obey equations of the form of (\ref{Newton}), where the gravitational potential $\phi$ is determined by the total energy density.  We will neglect baryon pressure, but consider both the normal and the quantum pressures of the superfluid. Hence, the equation of motion for baryons is
\be \label{NewtonB}
{\ddot{r}}_b^{i} \, = \, - \frac{\partial \phi}{\partial r_b^{i}} \, ,
\ee
while that of the superfluid dark matter is
\be \label{NewtonS}
{\ddot{r}}_{\mathrm{DM}}^{i} \, = \, - \frac{\partial \phi}{\partial r_{\mathrm{DM}}^{i}} + \frac{1}{\rho_{\mathrm{DM}}} \frac{\partial}{\partial r_{\mathrm{DM}}^{i}} (P + P_{QP}) \, ,
\ee
where $P$ and $P_{QP}$ are the normal and quantum pressure terms of the dark matter superfluid, respectively. 

The gravitational force at a distance $\mathbf{r}$ from the wake can be determined using conservation of energy inside the shell:
\be
	\rho\left(\mathbf{r}, t\right) d^{3}r = a^{3}\left(t\right) \bar{\rho}\left(t\right) d^{3}x,
\ee
where $\bar{\rho}$ is the background density.  This implies
\be
	\rho\left(\mathbf{r}, t\right) = \frac{a^{3}\left(t\right) \bar{\rho}\left(t\right)}{\left\vert\det\left(\partial r^{i}/\partial x^{j}\right)\right\vert} \simeq \bar{\rho}\left(t\right) \left(1 - \frac{\partial \psi^{i}}{\partial x^{i}}\right) \, .
\ee
Integrating the Poisson equation over a box about the wake leads to the result
\be \label{int-Poisson}
\frac{\partial\phi}{\partial r^{i}} \approx \frac{4 \pi G}{3} \left[\bar{\rho}_{b} \left(r_{b}^{i} - 3 a \psi_{b}^{i}\right) + \bar{\rho}_{\mathrm{DM}} \left(r_{\mathrm{DM}}^{i} - 3 a \psi_{\mathrm{DM}}^{i}\right)\right].
\ee

Since the initial velocity kick on the particles points in the $x$-direction, the only non-null component of the perturbation $\mat{\psi}$ is $\psi^{x} \equiv \psi$; and since $\psi$ only depends on $x$, $\mathbf{x}$ can be replaced by $x$. 
Making use of (\ref{int-Poisson}) and inserting the expansions (\ref{exp}), we find that the Zel'dovich equations for dark matter and baryons, taking into account the pressure terms, are
\begin{widetext}
\begin{align}
	\ddot{\psi}_{b} + 2 \frac{\dot{a}}{a} \dot{\psi}_{b} + \left(1 + 2 \frac{\bar{\rho}_{b}}{\bar{\rho}_{m}}\right) \frac{\ddot{a}}{a} \psi_{b} &= \frac{\bar{\rho}_{\mathrm{DM}}}{\bar{\rho}_{m}} \frac{\ddot{a}}{a} \left(x_{\mathrm{DM}} - x_{b} - 2 \psi_{\mathrm{DM}}\right),\\
	\ddot{\psi}_{\mathrm{DM}} + 2 \frac{\dot{a}}{a} \dot{\psi}_{\mathrm{DM}} + \left(1 + 2 \frac{\bar{\rho}_{\mathrm{DM}}}{\bar{\rho}_{m}}\right) \frac{\ddot{a}}{a} \psi_{\mathrm{DM}} &= \frac{\bar{\rho}_{b}}{\bar{\rho}_{m}} \frac{\ddot{a}}{a} \left(x_{b} - x_{\mathrm{DM}} - 2 \psi_{b}\right) + \frac{1}{a \rho_{\mathrm{DM}}} \nabla\left(P + P_{\mathrm{QP}}\right),
\end{align}
where $\rho_{m} = \rho_{b} + \rho_{\mathrm{DM}}$ is the energy density of matter.
The quantum pressure term is given by \cite{pressure}
\begin{align}
	\nabla P_{\mathrm{QP}} = - \frac{1}{4 m^{2}} \left[\nabla\left(\nabla^{2} \rho_{\mathrm{DM}}\right) - 2 \frac{\nabla^{2} \rho_{\mathrm{DM}}}{\rho_{\mathrm{DM}}} \nabla \rho_{\mathrm{DM}} + \frac{\left(\nabla\rho_{\mathrm{DM}}\right)^{2}}{\rho_{\mathrm{DM}}^{2}} \nabla\rho_{\mathrm{DM}}\right],
\end{align}
or, in terms of $\psi$, 
\be
 \partial_{r} P_{\mathrm{QP}} = \frac{a \bar{\rho}_{\mathrm{DM}}}{4 m^{2}} \left[\partial_{r}^{4}\psi_{\mathrm{DM}} + 2 a \left(1 + a \partial_{r}\psi_{\mathrm{DM}}\right) \partial_{r}^{3} \psi_{\mathrm{DM}} \partial_{r}^{2}\psi_{\mathrm{DM}} + a^{2} \left(1 + 2 a \partial_{r}\psi_{\mathrm{DM}}\right) \left(\partial_{r}^{2}\psi_{\mathrm{DM}}\right)^{2} \partial_{r}^{2}\psi_{\mathrm{DM}}\right],
 \ee
 \end{widetext}
 where $m$ is the mass of a superfluid quantum. This, in turn, implies
 \be
 \frac{\partial_{r} P_{\mathrm{QP}}}{a \rho_{\mathrm{DM}}} = \frac{1}{4 m^{2}} \partial_{r}^{4}\psi_{\mathrm{DM}},
\ee
up to first order in $\psi$.

The normal pressure term is determined by the equation of state, which at zero temperature (a good approximation since we are considering accretion during the matter-dominated period of the evolution of the universe) is given by
\begin{align} \label{pressure}
	P = \frac{\rho_{\mathrm{DM}}^{3}}{12 \Lambda^{2} m^{6}},
\end{align}
where the scale $\Lambda$ is related to the critical MONDian acceleration $a_0$ via 
$\Lambda \sim \sqrt{a_0 M_{\mathrm{Pl}}} \sim {\rm{meV}}$. Hence, making use of
$\rho_{\mathrm{DM}} \simeq \bar{\rho}_{\mathrm{DM}} (1 - a \partial_{r}\psi_{\mathrm{DM}})$ and the fact that to leading order $\psi_{\mathrm{DM}}$ is independent of $r$, we obtain
\ba
 \frac{\partial_{r}P}{a \rho_{\mathrm{DM}}} \, &=& \, \frac{\rho_{\mathrm{DM}}}{4 \Lambda^{2} m^{6} a} \partial_{r}\rho_{\mathrm{DM}} \\ 
 &\simeq& \, - \frac{\bar{\rho}_{\mathrm{DM}}^{2}}{4 \Lambda^{2} m^{6}} \partial_{r}^{2}\psi_{\mathrm{DM}} . \nonumber
\ea
%
%\begin{align}
%	\nonumber \frac{\partial_{r}P}{a \rho_{\mathrm{DM}}} &= \frac{\rho_{\mathrm{DM}}}{4 \Lambda^{2} m^{6} a} \partial_{r}\rho_{\mathrm{DM}} = - \frac{\bar{\rho}_{\mathrm{DM}}^{2}}{4 \Lambda^{2} m^{6}} \left(1 - a \partial_{r}\psi_{\mathrm{DM}}\right) \partial_{r}^{2}\psi_{\mathrm{DM}}\\
%	&\simeq - \frac{\bar{\rho}_{\mathrm{DM}}^{2}}{4 \Lambda^{2} m^{6}} \partial_{r}^{2}\psi_{\mathrm{DM}}.
%\end{align}

Combining the above equations, we find
\begin{widetext}
\begin{align}
	\ddot{\psi}_{b} + \frac{4}{3 t} \dot{\psi}_{b} - \frac{2}{9 t^{2}} \left(1 + 2 \epsilon_{1}\right) \psi_{b} &= - \frac{2}{9 t^{2}} \left(1 - \epsilon_{1}\right) \left(x_{\mathrm{DM}} - x_{b} - 2 \psi_{\mathrm{DM}}\right), \label{baryoneq2}\\
	\ddot{\psi}_{\mathrm{DM}} + \frac{4}{3 t} \dot{\psi}_{\mathrm{DM}} - \frac{2}{9 t^{2}} \left(3 - 2 \epsilon_{1}\right) \psi_{\mathrm{DM}} &= - \frac{2 \epsilon_{1}}{9 t^{2}} \left(x_{b} - x_{\mathrm{DM}} - 2 \psi_{b}\right) + \frac{1}{4 m^{2}} \partial_{r}^{4}\psi_{\mathrm{DM}} - \frac{\bar{\rho}_{\mathrm{DM}}^{2}}{4 \Lambda^{2} m^{6}} \partial_{r}^{2}\psi_{\mathrm{DM}}, \label{DMeq2}
\end{align}
\end{widetext}
where we expressed the density fractions in terms of $\epsilon_{1}$,
\be
	\epsilon_{1} \equiv \frac{\bar{\rho}_{b}}{\bar{\rho}_{m}} 	\,\,\,
	\rightarrow \,\,\, \frac{\bar{\rho}_{\mathrm{DM}}}{\bar{\rho}_{m}} = 1 - \epsilon_{1}.
\ee
Here, $\bar{\rho}_m$, $\bar{\rho}_{\mathrm{DM}}$ and $\bar{\rho}_b$ indicate the background densities of matter, dark matter and baryons, respectively. Note that the sign in the last term in (\ref{DMeq2}) reflects the fact that the normal pressure force is in the outward direction, i.e. it will lead to an increase in $\psi_{\mathrm{DM}}$ if $\psi_{\mathrm{DM}}$ is negative.

In the case of pressureless dark matter, then (as long as we neglect the baryon pressure) dark matter and baryons evolve in parallel. This means that in the above equations we can set $\epsilon_1 = 0$, and the equation of motion becomes
\be \label{CDMeq}
\ddot{\psi}_{a} + \frac{4}{3 t} \dot{\psi}_{a} - \frac{2}{3 t^{2}} \psi_{a} \,  = \, 0 \, ,
\ee
which has the dominant mode solution $\psi_{a}(t) \,  \sim \, t^{2/3} $,  i.e. $|\psi_{a}(t)|$ grows in proportion to the scale factor.
 
Now, let us look at the coefficients of the pressure terms. In superfluid dark matter models, we have $m \, \sim \,  {\rm{eV}}$ and $\Lambda \, \sim \, {\rm{meV}}$. The critical density today is $\rho_{c,0} \, \simeq \, 3.7 \times 10^{-11} {\rm{eV}}^{4}$, which gives $\bar{\rho}_{\mathrm{DM}} \simeq 10^{-11}$~eV$^{4}$, and we find
\begin{align} \label{epsilondef}
	\frac{\bar{\rho}_{\mathrm{DM}}^{2}}{4 \Lambda^{2} m^{6}} &\sim 10^{-17} \equiv \epsilon_{2},	&	\frac{1}{4 m^{2}} &\sim 10^{-1} \mathrm{eV}^{-2},
\end{align}
with $\epsilon_{2} \ll 1$.  Note that $\epsilon_2$ depends on time and the value quoted here is the value of $\epsilon_2$ evaluated at the present time. 

Thus,
\begin{widetext}
\begin{align}
	\ddot{\psi}_{b} + \frac{4}{3 t} \dot{\psi}_{b} - \frac{2}{9 t^{2}} \left(1 + 2 \epsilon_{1}\right) \psi_{b} &= - \frac{2}{9 t^{2}} \left(1 - \epsilon_{1}\right) \left(x_{\mathrm{DM}} - x_{b} - 2 \psi_{\mathrm{DM}}\right),\\
	\ddot{\psi}_{\mathrm{DM}} + \frac{4}{3 t} \dot{\psi}_{\mathrm{DM}} - \frac{2}{9 t^{2}} \left(3 - 2 \epsilon_{1}\right) \psi_{\mathrm{DM}} &= - \frac{2 \epsilon_{1}}{9 t^{2}} \left(x_{b} - x_{\mathrm{DM}} - 2 \psi_{b}\right) - \epsilon_{2} \partial_{r}^{2}\psi_{\mathrm{DM}} + \frac{1}{4 m^{2}} \partial_{r}^{4}\psi_{\mathrm{DM}}.
\end{align}
\end{widetext}
Estimating $\partial_{r}\psi_{\mathrm{DM}} \sim \psi_{\mathrm{DM}}/r_{\mathrm{DM}}$, we have that
\be
\partial_{r}\psi_{\mathrm{DM}} \sim \frac{\psi_{\mathrm{DM}}}{r_{\mathrm{DM}}} \, ,
\ee
\ba	
\partial_{r}^{2}\psi_{\mathrm{DM}} &\sim& \frac{1}{r_{\mathrm{DM}}} \left(\partial_{r}\psi_{\mathrm{DM}} - \frac{\psi_{\mathrm{DM}}}{r_{\mathrm{DM}}}\right) \sim- \frac{\psi_{\mathrm{DM}}}{r_{\mathrm{DM}}^{2}},
\ea
\ba
	\partial_{r}^{3}\psi_{\mathrm{DM}} &\sim& \frac{1}{r_{\mathrm{DM}}} \left(\partial_{r}^{2}\psi_{\mathrm{DM}} - \frac{2 \partial_{r}\psi_{\mathrm{DM}}}{r_{\mathrm{DM}}} + \frac{2 \psi_{\mathrm{DM}}}{r_{\mathrm{DM}}^{2}}\right) \nonumber \\
	 &\sim& \frac{2 \psi_{\mathrm{DM}}}{r_{\mathrm{DM}}^{3}},
\ea
\ba
	\partial_{r}^{4}\psi_{\mathrm{DM}} &\sim& \frac{1}{r_{\mathrm{DM}}} \left(\partial_{r}^{3}\psi_{\mathrm{DM}} - \frac{3 \partial_{r}^{2}\psi_{\mathrm{DM}}}{r_{\mathrm{DM}}} + \frac{6 \partial_{r}\psi_{\mathrm{DM}}}{r_{\mathrm{DM}}^{2}} - \frac{6 \psi_{\mathrm{DM}}}{r_{\mathrm{DM}}^{3}}\right) \nonumber \\
	&\sim& - \frac{6 \psi_{\mathrm{DM}}}{r_{\mathrm{DM}}^{4}},
\ea
so that the equations become
\begin{widetext}
\begin{align}
	\ddot{\psi}_{b} &+ \frac{4}{3 t} \dot{\psi}_{b} - \frac{2}{9 t^{2}} \left(1 + 2 \epsilon_{1}\right) \psi_{b} \, = \, - \frac{2}{9 t^{2}} \left(1 - \epsilon_{1}\right) \left(x_{\mathrm{DM}} - x_{b} - 2 \psi_{\mathrm{DM}}\right),\\
	\ddot{\psi}_{\mathrm{DM}} &+ \frac{4}{3 t} \dot{\psi}_{\mathrm{DM}} - \frac{2}{9 t^{2}} \left(3 - 2 \epsilon_{1}\right) \psi_{\mathrm{DM}} - \epsilon_{2} \frac{\psi_{\mathrm{DM}}}{r_{\mathrm{DM}}^{2}} + \frac{3 \psi_{\mathrm{DM}}}{2 m^{2} r_{\mathrm{DM}}^{4}} \, = \, - \frac{2 \epsilon_{1}}{9 t^{2}} \left(x_{b} - x_{\mathrm{DM}} - 2 \psi_{b}\right),
\end{align}
or, in terms of $x$,
\begin{align}
	\ddot{\psi}_{b} &+ \frac{4}{3 t} \dot{\psi}_{b} - \frac{2}{9 t^{2}} \left(1 + 2 \epsilon_{1}\right) \psi_{b} \, = \, - \frac{2}{9 t^{2}} \left(1 - \epsilon_{1}\right) \left(x_{\mathrm{DM}} - x_{b} - 2 \psi_{\mathrm{DM}}\right), \label{beq3}\\
	\ddot{\psi}_{\mathrm{DM}} &+ \frac{4}{3 t} \dot{\psi}_{\mathrm{DM}} - \frac{2}{9 t^{2}} \left(3 - 2 \epsilon_{1}\right) \psi_{\mathrm{DM}} 
	+ \left(\frac{t_{i}}{t}\right)^{4/3} \frac{\epsilon_{2} \psi_{\mathrm{DM}}}{a^{2}\left(t_{i}\right) x_{\mathrm{DM}}^{2}} + \left(\frac{t_{i}}{t}\right)^{8/3} \frac{3 \psi_{\mathrm{DM}}}{2 m^{2} a^{4}\left(t_{i}\right) x_{\mathrm{DM}}^{4}} \simeq - \frac{2 \epsilon_{1}}{9 t^{2}} \left(x_{b} - x_{\mathrm{DM}} - 2 \psi_{b}\right). \label{DMeq3}
\end{align}
\end{widetext}

We now have the equations required to determine the widths of the dark matter and baryonic components of the wake.  In the following we will study the effects of superfluid quantum pressure, normal pressure, and coupling to baryons on the string wake.

 \section{Effect of Quantum Pressure}
 \label{sec3}
 
Before studying the effects of quantum pressure on the dark matter wake we will review the determination of the thickness of a pure cold dark matter wake. At time $t$, the width of the wake is given by the value of $x$ for which ${\dot{r}} = 0$.  Solving the equation of motion (\ref{CDMeq}) with initial conditions given by (\ref{IC}) yields the solution
\be \label{CDMsol}
 \psi_{\mathrm{DM}}(t) \equiv \psi^{0}(t) \, \simeq \, - \frac{3 u_i t_i}{5 a(t_i)} \left[\left(\frac{t}{t_i} \right)^{2/3}
 - \left( \frac{t_i}{t} \right) \right] \, .
\ee
Note that for the Wronskian analysis which follows we need to keep both the growing and the decaying modes. The shell turning around at time $t$ then has the value (see (\ref{width}))
\be \label{CDMwidth}
x_{\mathrm{DM}}(t) \, \simeq \, \frac{6 u_i t_0}{5 \sqrt{a(t_i)}} a(t) \, 
\ee
(here we can focus on the growing mode exclusively). After turnaround the wake virializes at a width of about 1/4 of that value. Note that the comoving width of the wake increases linearly with $a(t)$ as expected from linear perturbations theory, while the physical width grows as $a(t)^2$.
 
 Let us now turn on the effects of the quantum pressure term, while still neglecting the normal pressure and the baryons (i.e. setting $\epsilon_1 = \epsilon_2 = 0$). In this case, the equation for $\psi_{\mathrm{DM}}$ becomes\footnote{In the following we consider the evolution of baryons and dark matter starting on the same shell, i.e. $x_{\mathrm{DM}} = x_b \equiv x$.}
 \be \label{EoM1}
 \ddot{\psi}_{\mathrm{DM}} + \frac{4}{3 t} \dot{\psi}_{\mathrm{DM}} - \frac{2}{3 t^{2}}  \psi_{\mathrm{DM}} \, = \, - \frac{3 \psi_{\mathrm{DM}}}{2 m^{2} a^{4}\left(t_{i}\right) x_{\mathrm{DM}}^{4}} \left(\frac{t_{i}}{t}\right)^{8/3}  \, .
 \ee
 We see that since $\psi_{\mathrm{DM}} < 0$, the extra force is repulsive and will counteract gravitational accretion.

In general, the equation of motion (\ref{EoM1}) is difficult to solve exactly. However, for the values of $G\mu$ in the range where strings leave behind interesting and potentially observable signatures in cosmological observations\footnote{The robust upper bound on $G\mu$ from the non-observation of effects of long strings on the angular power spectrum of CMB anisotropies \cite{CMBbound} is $G\mu < 10^{-7}$, and for values of $G\mu$ in the range of $10^{-10}$ interesting gravitational wave signals in pulsar timing arrays \cite{PTA} and predicted, and strings with a similar value of the string tension would have a large impact on the abundance of high redshift galaxies \cite{JH1}, and might also explain the abundance of supermassive black holes observed at high redshifts \cite{JH2} (see also \cite{Richhild} for some early work).} the source term on the right hand side of the equation is small compared to the other terms, and hence we can make use of the Born approximation method to solve the equation perturbatively.

To check the above statement, we take the magnitude of $x_{\mathrm{DM}}$ to correspond to the width of a wake created at time $t_i$.  In this case, the source term in (\ref{EoM1}) is smaller than the last term on the left hand side of the equation (evaluated at the time $t_{\mathrm{eq}}$) provided that
\be
m t_{\mathrm{eq}} \, \gg (G \mu)^{-2} \, .
\ee
Inserting the value $m \sim {\rm{eV}}$ we find that the above inequality is obeyed provided that $G\mu > 10^{-13}$.

Defining the dimensionless factor
\begin{align}
	\alpha(x_{\mathrm{DM}}) &\equiv \frac{3 t_{i}^{2}}{2 m^{2} a^{4}\left(t_{i}\right) x_{\mathrm{DM}}^{4}},
\end{align}
we can rewrite our equation as
\be \label{EoM2}
	\ddot{\psi}_{\mathrm{DM}} + \frac{4}{3 t} \dot{\psi}_{\mathrm{DM}} + \left[\frac{\alpha}{t_{i}^{2}} \left(\frac{t_{i}}{t}\right)^{8/3} - \frac{2}{3 t^{2}}\right] \psi_{\mathrm{DM}} \, = \, 0.
\ee

Now, for $\alpha \ll 1$, we can make use of the Born approximation to find an approximate solution.  We make the ansatz
\be
\psi_{\mathrm{DM}}^{\alpha} \, \equiv \, \psi^{0} + \delta\psi_{\mathrm{DM}}^{\alpha} \, ,
\ee
where $\psi^{0}$ is the solution for pressureless dark matter (i.e. for $\alpha = 0$), and $\delta\psi_{\mathrm{DM}}^{\alpha}$ is the correction term which is, to leading order,  linear in $\alpha$. By inserting this ansatz into (\ref{EoM2}) and working to linear order in $\alpha$, the equation for the correction term is obtained by using the $\alpha = 0$ solution in the source term:
\ba \label{EoM3}
	\ddot{\delta\psi}_{\mathrm{DM}}^{\alpha} &+& \frac{4}{3 t} \dot{\delta\psi}_{\mathrm{DM}}^{\alpha} - \frac{2}{3 t^{2}} \delta\psi_{\mathrm{DM}}^{\alpha} = - \frac{\alpha}{t_{i}^{2}} \left(\frac{t_{i}}{t}\right)^{8/3} \psi^{0} \nonumber \\
	&=& \frac{3 \alpha u_{i}}{5 a\left(t_{i}\right) t_{i}} \left[\left(\frac{t_{i}}{t}\right)^{2} - \left(\frac{t_{i}}{t}\right)^{11/3}\right] \, ,
\ea
where it is important to keep both terms in $\psi^{0}$.This inhomogeneous equation can be solved using the Green's function method. First, we need to compute the Wronskian of the system, which is obtained from the solutions $y_1$ and $y_2$ of the homogeneous equation,
\begin{align} 
y_1(t) \, &= \, \left(\frac{t}{t_i}\right)^{2/3},	&	y_2(t) \, &= \, \left(\frac{t_i}{t}\right)^{-1},  \label{fundsol1}
\end{align}
as
\begin{align} \label{Wronskian}
	W\left(t\right) \, = \, y_{1} \dot{y}_{2} - y_{2} \dot{y}_{1} \, = \, - \frac{5}{3 t_{i}} \left(\frac{t_{i}}{t}\right)^{4/3} \, .
\end{align}
The solution of (\ref{EoM3}) with initial conditions given by (\ref{IC}) is given by adding to the solution of the homogeneous equation obeying the given initial conditions the particular solution with vanishing initial data
\be \label{GF}
\delta\psi_{\mathrm{DM}}^{\alpha} \, = \, - y_{1} \int_{t_{i}}^{t} \frac{y_{2} s}{W} dt^{\prime} + y_{2} \int_{t_{i}}^{t} \frac{y_{1} s}{W} dt^{\prime} \, ,
\ee
where $s(t)$ is the source term, the right hand side of (\ref{EoM3}).  Performing the integral and keeping only the dominant term for $t \gg t_i$ gives the result
\be \label{bdm}
\delta\psi_{\mathrm{DM}}^{\alpha}\left(t, x_{\mathrm{DM}}\right) \, \simeq \,  
\frac{27 u_i t_i}{70 a(t_i)} \alpha\left(x_{\mathrm{DM}}\right) \left( \frac{t}{t_i} \right)^{2/3},
\ee
where both sides of the equation depend on the value $x_{\mathrm{DM}}$ of the shell we are considering. The full solution for $\alpha \ll 1$ is thus
\be
\psi_{\mathrm{DM}}^{\alpha}\left(t, x_{\mathrm{DM}}\right) \, = \, - \frac{3 u_i t_i}{5 a(t_i)} \left( \frac{t}{t_i} \right)^{2/3}  
\left[ 1 - \frac{9}{14} \alpha\left(x_{\mathrm{DM}}\right)  \right]
\ee
up to first order in $\alpha$. As expected, the quantum pressure term partially counteracts the gravitational force and hence causes the shells to collapse more slowly.
 
\begin{figure}[t]
	\centering
	\includegraphics[scale=.5]{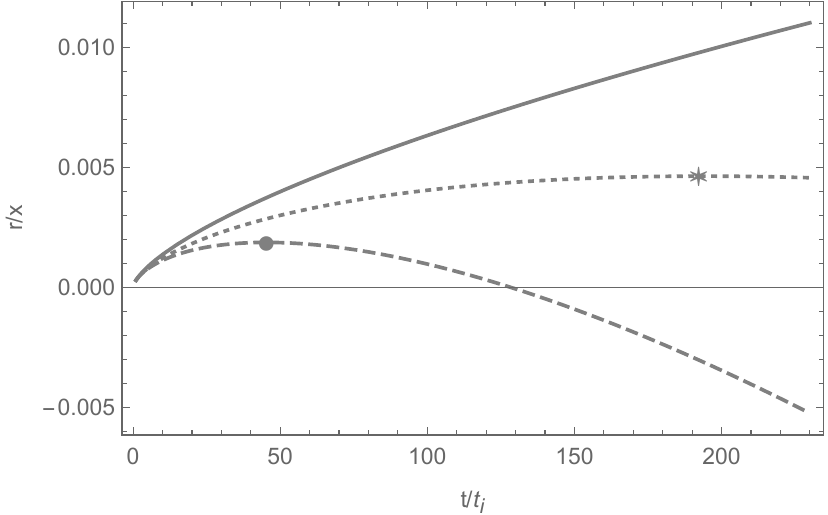}
	\caption{Evolution of a mass shell as a function of time (horizontal axis). The vertical axis gives the physical height of the shell above the center of the wake.  The solid curve shows the increase in the physical height without any gravitational attraction, the other two curves include the effects of the wake, the dashed curve showing the physical  height of the shell in the case of pure cold dark matter, and the dotted curve giving the physical height when the quantum pressure of the superfluid dark matter is included.  The turnaround times (times when the shell starts to collapse towards the center) are marked with a circle (for the pure cold cold matter case) and a star (in the case of superfluid dark matter). We chose the value of $x = x_{\mathrm{crit}}$ such that the effects of the quantum pressure are visible. To make this shell collapse for a reasonable value of $t/t_i$ we chose the value $t_i m = 10^{10}$.  The other parameter values chosen are $\Lambda = 10^{-3} {\rm{eV}}$, $G\mu = 10^{-7}$, $t_i = t_{\mathrm{eq}}$ (the time of equal matter and radiation which corresponds to redshift $z_{\mathrm{eq}} = 3400$), $v_s \gamma_s = 3^{-1/2}$, $\Omega_m = 0.315$ and $\Omega_b = 0.05$. Note that the Zel'dovich approximation breaks down after turnaround.}
	\label{f01.png}
\end{figure}

In Fig.~\eqref{f01.png} we show the evolution of the mass shell without any source (solid curve), for pure cold dark matter in the presence of the wake source (dashed curve), and when including the effects of dark matter quantum pressure (dotted curve). The horizontal axis is time, the vertical axis is the physical height (after dividing by $x$).  The effect of the quantum pressure at decreasing the rate of accretion and making the wakes thinner is manifest -- in the presence of the quantum pressure, a fixed initial comoving scale turns around later, and thus at a fixed time the comoving scale which is turning around is smaller, and hence the wake is thinner.
 
\begin{figure}[t]
	\centering
	\includegraphics[scale=.5]{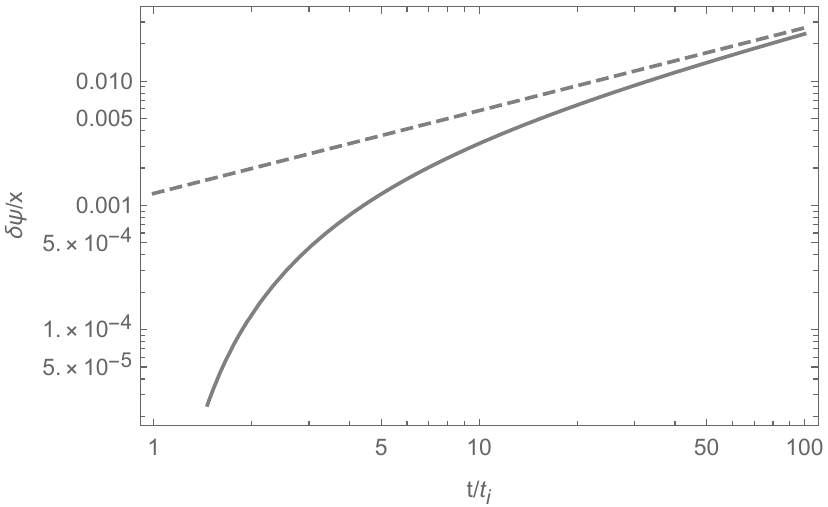}
	\caption{Comparison between the approximate result for $\delta\psi_{\rm{DM}}^{\alpha}$ (dashed curve) with the full result (solid line), for the value $x = 10^2 x_{\mathrm{crit}}$. The horizontal axis is time (in units of $t_i$), and the vertical axis is the value of $\delta\psi_{\rm{DM}}^{\alpha}/x$. The values of the parameters used (in addition to the values of $m = 1{\rm eV}$ and $\Lambda = 10^{-3} {\rm eV}$ mentioned in the main text) are $G\mu = 10^{-7}$, $t_i = t_{\mathrm{eq}}$, $v_s \gamma_s = 3^{-1/2}$, $\Omega_m = 0.315$ and $\Omega_b = 0.05$.}
	\label{fig02}
\end{figure}

 In Fig.~\eqref{fig02}, we compare the solution for $\delta\psi_{\rm{DM}}^{\alpha}$ (solid curve) with the approximate result of (\ref{bdm}).  The plot is for the value $x_b = x_{\rm{DM}} = 10^{2} x_{\rm{crit}}$, where $x_{\rm{crit}}$ is the value of $x$ for which $\alpha = 1$:
 \be
 x_{\rm{crit}} \, = \, \left( \frac{3}{2} \right)^{1/4} \sqrt{\frac{t_i}{m}} \frac{1}{a(t_i)} \, .
 \ee
 As is apparent, the analytical approximation of (\ref{bdm}) (which only keeps the growing mode) is a good one at late times.
 
 The wake width is given by (\ref{width}) and yields
 \be
 x_{\mathrm{DM}}(t) \, = \frac{6 u_i t_i}{5} \frac{a(t)}{a(t_i)}  \left[ 1 - \frac{9}{14} \alpha(x_{\mathrm{DM}})  \right] \, .
 \ee
 We conclude that the presence of quantum pressure leads to a decrease in the wake width by a factor $F(x_{\mathrm{DM}})$,
 \be \label{factor}
 F(x_{\mathrm{DM}}) \, = \, 1 - \frac{9}{14} \alpha(x_{\mathrm{DM}}) \, .
 \ee
%%
%This yields the relative decrease of the wake width at time $t$, we can, to leading order in $\alpha$, insert the regular CDM relation $x(t)$ from (\ref{CDMwidth}).

Let us now turn to the induced baryon wake,  still setting $\epsilon_1 = 0$. The comoving displacement of the baryon fluid obeys the equation
\be \label{EoM4}
\ddot{\psi}_{b} + \frac{4}{3 t} \dot{\psi}_{b} - \frac{2}{9 t^{2}}  \psi_{b} \, = \,  \frac{4}{9 t^{2}}  \psi_{\mathrm{DM}} \, .
\ee
The effect of the quantum pressure is to generate a difference between the dark matter and baryon wakes. Hence, we expand
\be
\psi_b \, = \, \psi^{0} + \delta \psi_b \, ,
\ee
where $\delta \psi_b$ obeys the equation
\be \label{EoM5}
\ddot{\delta \psi}_{b} + \frac{4}{3 t} \dot{\delta \psi}_{b} - \frac{2}{9 t^{2}}  \delta \psi_{b} \, = \,  \frac{4}{9 t^{2}}  \delta \psi_{\mathrm{DM}} \, ,
\ee
with vanishing initial conditions at time $t_i$. This equation can again be solved using the Green's function method. 

\begin{figure}[t]
	\centering
	\includegraphics[scale=.5]{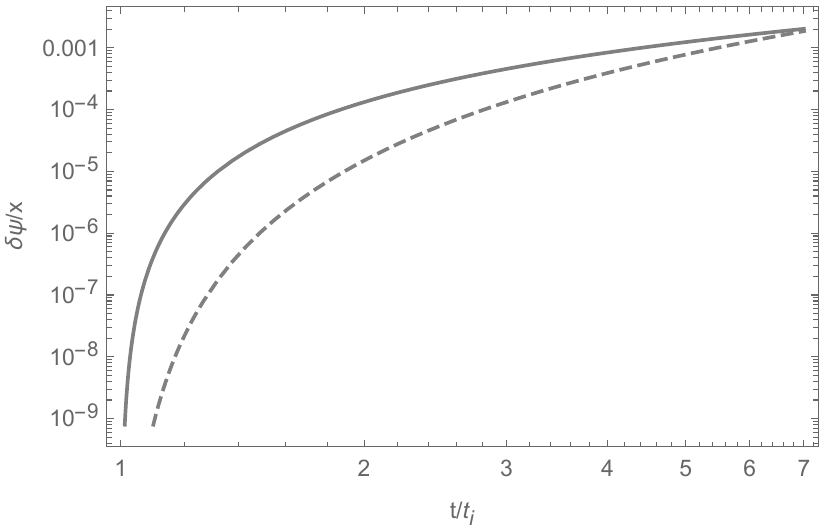}
	\caption{Comparison between the values of $\delta\psi$ (including quantum pressure effects) for baryons and dark matter, choosing $x = 10^2 x_{\rm{crit}}$. It is apparent that the effect of the quantum pressure on the baryon shells ($\delta \psi_b^{\alpha}$ -- dashed curve) is smaller than for the dark matter shells ($\delta\psi_{\rm{DM}}^{\alpha}$ -- solid curve), although the difference becomes negligible at late times. The horizontal axis is time (in units of $t_i$), the vertical axis is the values of $\delta\psi/x$. The parameter values are the same as those in Figure (\ref{fig02}).}
	\label{fig03}
\end{figure}

The fundamental solutions of the homogeneous equation are 
\begin{align}
	y_1(t) &= \left(\frac{t}{t_i}\right)^{1/3},	&	y_2(t) &= \left(\frac{t_i}{t}\right)^{2/3}, \label{fundsol2}
\end{align} 
and the resulting Wronskian is (see (\ref{Wronskian}))
\be \label{Wronskian2}
W(t) \, = \, - \frac{1}{t_i} \left(\frac{t_{i}}{t}\right)^{4/3} \  \, .
\ee
The source term on the right hand side of (\ref{EoM5}) is
\be
s(t) \, = \, \frac{6 \alpha u_i}{35 a(t_i) t_i} \left(\frac{t_i}{t}\right)^{4/3} \, .
\ee
Hence, making use of the Green's function formula (\ref{GF}), we find
\be \label{bwake}
\delta \psi_b(t) \, \simeq \, \frac{27 \alpha u_i t_i}{70 a(t_i)} \left( \frac{t}{t_i}\right)^{2/3} \, .
\ee
Note that the leading terms in $\delta \psi_{\mathrm{DM}}$ (\ref{bdm}) and $\delta \psi_b$ (\ref{bwake}) which we have computed are identical.  This implies that at late times the effect of dark matter quantum pressure is the same for baryons and dark matter. This is due to the fact that the effects of the quantum pressure tend to zero as the width of the wake increases. 
%Note that the leading term in $\delta \psi_{\mathrm{DM}}$ (\ref{bdm}) is larger by a factor of 2 compared to $\delta \psi_b$ (\ref{bwake}). 
If we take into account the subleading terms, we find that at early times $|\delta \psi_b| < |\delta \psi_{\mathrm{DM}}|$ and thus that the effects of quantum pressure are more important for the dark matter than for the baryons, and that the baryon wake is hence slightly larger than the dark matter wake.  The numerical results keeping all terms are shown in Fig.~\eqref{fig03}. For the complete equations see \cite{AlineThesis}.
 
\section{Effect of Normal Pressure}
\label{sec4}

We now turn to a discussion of the effects of the normal pressure of the dark matter, i.e. we now consider $\epsilon_2 \neq 0$. We will turn off the effects of quantum pressure. Since in the linear analysis both quantum and normal pressure act as source terms, their contributions to $\delta \psi_{\mathrm{DM}}$ and $\delta \psi_b$ add up, i.e.
\be
\delta \psi_{\mathrm{DM}}^{\epsilon_2, \alpha}\, = \, \delta \psi_{\mathrm{DM}}^{\epsilon_2} + \delta \psi_{\mathrm{DM}}^{\alpha} \, ,
\ee
where the superscripts $\alpha$ and $\epsilon_2$ stand for the contributions which come from the quantum and normal pressure terms alone. In the following we compute $\delta \psi_{\mathrm{DM}}^{\epsilon_2}$, and we drop the superscript for notational convenience.

The equation for $\delta \psi_{\mathrm{DM}}$ is
\be \label{EoM8}
\ddot{\delta \psi}_{\mathrm{DM}} + \frac{4}{3 t} \dot{\delta \psi}_{\mathrm{DM}} - \frac{2}{3 t^{2}}  \delta \psi_{\mathrm{DM}} \, = \, - \frac{\epsilon_2 \psi^{0}}{a^2(t_i) x_{\mathrm{DM}}^2} \left( \frac{t_i}{t} \right)^{4/3} \, ,
\ee
where $\psi^{0}$ is the unperturbed solution, the solution for pure cold dark matter given by (\ref{CDMsol}).

Extracting the time dependence of the background energy density which appears in the expression for $\epsilon_2$, we can write
\be
\epsilon_2(t) \, \equiv \, {\tilde{\epsilon}_2} \left( \frac{t_0}{t} \right)^4 ,
\ee
with ${\tilde{\epsilon}_2}$ given by the defining expression (\ref{epsilondef}) with the background density evaluated at the present time $t_0$.

The equation of motion (\ref{EoM8}) can again be solved by means of the Green's function method. The fundamental solutions and Wronskian are the same as the ones which appear in (\ref{fundsol1}) and (\ref{Wronskian}).  We find that the term which dominates at late times takes the form
\ba \label{result3}
\delta \psi_{\mathrm{DM}}(t) \, &\simeq& \, \frac{9 \tilde{\epsilon}_2 u_i t_0^4}{250 a^3(t_i) x_{\mathrm{DM}}^2 t_i} \left( \frac{t}{t_i} \right)^{2/3} \, \nonumber \\
&\equiv& \, \kappa \left( \frac{t}{t_i} \right)^{2/3},
\ea
where the constant $\kappa$ is defined by the last equality.

\begin{figure}[t]
	\centering
	\includegraphics[scale=.5]{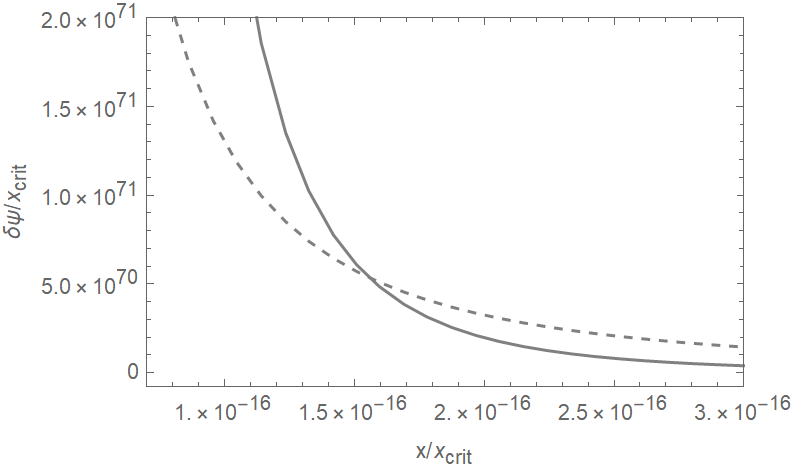}
	\caption{Comparison between the effects of quantum and normal pressure. The solid curve shows the change in the comoving height of the shell $\delta\psi_{\rm{DM}}^{\alpha}$ due to quantum pressure, the dashed curve shows the effects of the normal pressure $\delta \psi_{\mathrm{DM}}^{\epsilon_2}$, both as a function of the initial comoving shell height $x$ (horizontal axis) and evaluated at the time $t = 10 t_i$. The parameter values are the same as those in Fig.~(\ref{fig02}).  For these parameter values, the effects of the normal pressure dominate, and the crossover occurs for such a small value of $x$ such that the value of $\psi$ is non-physical (long after turnaround has occurred). }  
	\label{fig04}
\end{figure}

Comparing the effects of quantum pressure (see (\ref{bdm})) and normal pressure (\ref{result3}), we see that the quantum pressure effects are more important than those of the normal pressure for small distances (small values of $x_{\mathrm{DM}}$) and for wakes produced at later times (larger values of $t_i$ -- see Fig.~\eqref{fig04}). Both contributions scale with time in the same way.

The leading order equation for the fluctuation in the baryon shell position is
\ba \label{EoM9}
\ddot{\delta \psi}_b + \frac{4}{3 t} \dot{\delta \psi}_b - \frac{2}{9 t^{2}}  \delta \psi_b \, &=& \, \frac{4}{9 t^2} \delta \psi_{\mathrm{DM}} \\
&=& \, \frac{4 \kappa}{9 t_i^{2}} \left( \frac{t_i}{t} \right)^{4/3} \, ,  \nonumber
\ea
which, using the Green's function method and the fundamental solutions and Wronskian from (\ref{fundsol2}) and (\ref{Wronskian2}), leads to the solution
\be
\delta \psi_b\left(t\right) \, \simeq \, \kappa  \left( \frac{t}{t_i} \right)^{2/3} \, ,
\ee
the same result as for the dark matter fluctuation. Once again, at sub-leading order there is a difference between $\delta \psi_{\mathrm{DM}}$ and $\delta \psi_b$.

Let us now compare the magnitude of the effects of quantum pressure and normal pressure of the superfluid on the position of an accreting mass shell. The ratio of $\delta \psi_{\mathrm{DM}}^{\alpha}$ and $\delta \psi_{\mathrm{DM}}^{\epsilon_{2}}$ gives
\be
\frac{\delta \psi_{\mathrm{DM}}^{\alpha}}{\delta \psi_{\mathrm{DM}}^{\epsilon_2}}
\, = \, \frac{450}{7} \left( \frac{t_i}{t_0} \right)^4 \frac{\Lambda^2 m^4}{\bar{\rho}^2_{\mathrm{DM}}(t_0)}
\frac{1}{a^2(t_i) x_{\mathrm{DM}}^2} \, .
\ee
The quantum pressure effects dominate on short distance scales, since the quantum pressure force scales as $x_{\mathrm{DM}}^{-4}$ compared to the normal pressure force, which scales at $x_{\mathrm{DM}}^{-2}$. Since the normal pressure is a rapidly decreasing function of time,  the quantum pressure is more important for wakes which form late compared to wakes which form early. 

The initial shell height $a(t_i) \:x$ below which quantum pressure dominates obviously depends on the values of the parameters $m$ and $\Lambda$ of the superfluid dark matter model. For the values $m = 1 {\rm{eV}}$ and $\Lambda = 10^{-3} {\rm{eV}}$, we find that the quantum pressure only dominates on very small scales,
\be
a(t_i) \:x \, < \, \left( \frac{t_i}{t_0} \right)^2 2 \times 10^5 {\rm{cm}} \, .
\ee
However, it is easy to imagine parameter values for which the quantum pressure is important also on astrophysical scales.

\section{Effects of Baryon-Dark Matter Coupling}
\label{sec5}

If we turn off both pressure terms (i.e.  set $\alpha = \epsilon_2 = 0$), then the dark matter and baryon shells remain coincident, since they then have the same equation of state. This can also be seen explicitly from our basic equations (\ref{beq3}) and (\ref{DMeq3}): for $\alpha = \epsilon_2 = 0$, the ansatz $\psi_{\mathrm{DM}}(t) = \psi_b(t) = \psi^0(t)$ obeys the equations. Hence, the effects of $\epsilon_1$ only arise when $\alpha$ and/or $\epsilon_2$ are non-vanishing, and the effects are proportional to $\alpha \epsilon_1$ and $\epsilon_2 \epsilon_1$, respectively.

To study the effect of the coupling between dark matter and baryons, we start with the  ansatz
\ba
\psi_{\mathrm{DM}} \, &=& \, \psi^0 + \delta \psi_{\mathrm{DM}}^{\alpha} + \delta \psi_{\mathrm{DM}}^{\epsilon_2} + \delta \psi_{\mathrm{DM}}^{\epsilon_1}, \nonumber \\
\psi_b \, &=& \, \psi^0  + \delta \psi_b^{\alpha} + \delta \psi_b^{\epsilon_2} + \delta \psi_b^{\epsilon_1}, \, 
\ea
where $\psi^0$ is the pure cold dark matter solution,  $\delta \psi_{\mathrm{DM}}^{\alpha}$ and $\delta \psi_b^{\alpha}$ are the correction terms due to the quantum pressure (computed in Section (\ref{sec3})), $\delta \psi_{\mathrm{DM}}^{\epsilon_2}$ and $\delta \psi_b^{\epsilon_2}$ are those due to the normal pressure (computed in Section (\ref{sec4})), and $\delta \psi_{\mathrm{DM}}^{\epsilon_{1}}$ and $\delta \psi_b^{\epsilon_{1}}$ are the new correction terms due to the coupling between baryons and dark matter to be analysed in this section. In the remainder of this section, we shall drop the superscript $\epsilon_{1}$ for conciseness. After inserting this ansatz into the basic equations (\ref{beq3}) and (\ref{DMeq3}) we find, to leading order in $\alpha + \epsilon_2$, that
\be
\ddot{\delta \psi}_{\mathrm{DM}} + \frac{4}{3 t} \dot{\delta \psi}_{\mathrm{DM}} - \frac{2}{3 t^2} \delta \psi_{\mathrm{DM}} \, = \, - \frac{4 \epsilon_1}{9 t^2} \bigl( \psi_{\mathrm{DM}} - \psi_b \bigr), \,
\ee
where on the right hand side of this equation $\psi_{\mathrm{DM}}$ and $\psi_b$ are the expressions for the dark matter and baryon shell positions which include the leading order quantum and normal pressure contributions. The corresponding equation for the baryon shell fluctuation is
\be
\ddot{\delta \psi}_b + \frac{4}{3 t} \dot{\delta \psi}_b - \frac{2}{9 t^2} \delta \psi_b \, = \, - \frac{4 \epsilon_1}{9 t^2} \bigl( \psi_{\mathrm{DM}} - \psi_b \bigr) \, .
\ee

\begin{figure}[t]
	\centering
	\includegraphics[scale=.5]{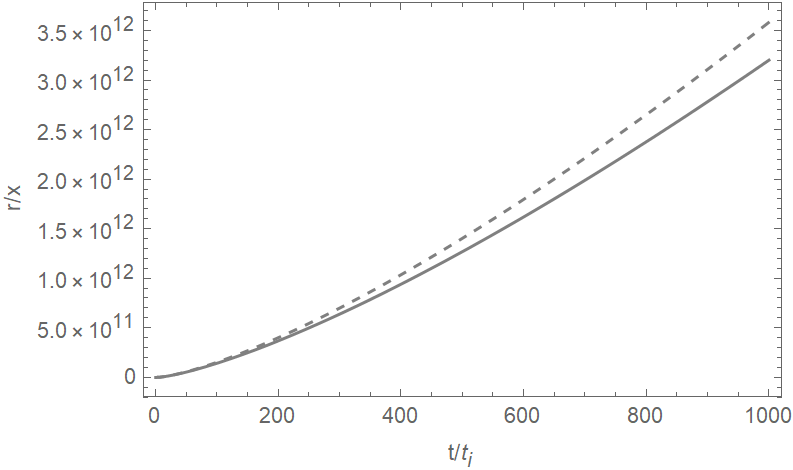}
	\caption{Effect of the coupling between dark matter and baryons: in the presence of such a coupling the value of $\delta\psi$ for the dark matter (solid curve) is smaller than without the coupling (dashed curve). This implies that the wake thickness is slightly larger in the presence of the coupling.The parameter values are the same as those in Fig.~(\ref{fig02}), except that $x = x_{\mathrm{crit}}$ (for significantly larger values of $x$ the difference between the two curves is negligible).}
	\label{fig05}
\end{figure}

As expected, it is the difference in $\psi_{\mathrm{DM}}$ and $\psi_b$ induced by the quantum and normal pressure effects which seeds the effects of the coupling $\epsilon_1$.  These effects can be computed explicitly using the same Green's function method applied in previous sections. Here, we will only discuss the qualitative aspects of the result. We have seen that the leading late time effects of both quantum and normal pressure on $\psi_{\mathrm{DM}}$ and $\psi_b$ are identical and hence do not lead to any effects of $\epsilon_1$. Beyond leading order, we have seen that both quantum and normal pressure effects lead to corrections to $\psi_{\mathrm{DM}}$ which are larger than the corrections to $\psi_b$.  From the above equations, we see that this induces a negative contribution to both $\delta \psi_{\mathrm{DM}}$ and $\delta \psi_b$. Due to the coupling between the baryons and the dark matter, the dark matter shells will be collapsing slightly faster.  This leads to the conclusion that the effects of the coupling will render the dark matter wake slightly wider than it would be in the absence of the coupling. The result is illustrated in Fig.~\eqref{fig05}, which shows the result obtained by solving the equations (see \cite{AlineThesis} for more details).

\section{Conclusions and Discussion} \label{conclusion}

 We have studied cosmic string wake formation in a set-up in which the dark matter is a superfluid.  We considered the effects of both quantum and normal pressures of the superfluid. The effects of both of these pressures counteracts the attractive force of gravity and causes mass shells to collapse more slowly. A given mass shell will have a comoving displacement which is smaller than what it would have been in the absence of pressure. This causes the wake to be thinner than it would be without the effects of pressure.

We have shown that the effects of quantum pressure dominate on small scales, and that they are more important for wakes seeded by string segments at late times $t_i$. This is due to the fact that the quantum pressure forces the wake to grow faster as a function of decreasing distance, and due to the fact that the normal pressure forces are rapidly diluted by the cosmological expansion. 

For the value of the mass $m$ which is a good fit to MOND, the effects of normal and quantum pressure forces are negligible on scales of cosmological interest. However, for lighter masses, or for wakes in the very early universe, the pressure forces could have important implications.

Note that by setting $\alpha = 0$ our analysis yields the effects to the string wake which normal pressure induces in any scenario in which the pressure of the dark matter cannot be neglected and can be described by (\ref{pressure}).

Our analysis has been to leading order in the strength of the pressure terms, and the gravitational coupling between baryons and dark matter. We have neglected the effects of shell crossing. This implies that our analysis cannot describe the internal structure of a wake, but only its overall size. On the other hand, it is the overall size and mass per unit area of the wake which is responsible for its gravitational effects.

\section*{Acknowledgement}

We wish to thank Heliudson Bernardo for many useful discussions. This research is supported in part by funds from NSERC and from the Canada Research Chair program.

\end{document}